\documentclass{aa}  

\usepackage{graphicx}
\usepackage{subfigure}
\usepackage{subfloat}
\usepackage{float}
\usepackage[version=3]{mhchem}
\usepackage{booktabs}
\usepackage{pdflscape}
\usepackage{supertabular}
\usepackage{longtable,lscape}
\usepackage{txfonts}
\begin{document} 

\title{Probing the protoplanetary disk gas surface density distribution with $^{13}$CO emission}


\author{A. Miotello\inst{\ref{inst:ESO},\ref{inst:leiden}} \and
	    S. Facchini\inst{\ref{inst:mpe}} \and
            E. F. van Dishoeck\inst{\ref{inst:leiden},\ref{inst:mpe}}\and 
	    S. Bruderer }

\institute{
European Southern Observatory, Karl-Schwarzschild-Str 2, D-85748 Garching, Germany\label{inst:ESO}\and
Leiden Observatory, Leiden University, Niels Bohrweg 2, NL-2333 CA Leiden, The Netherlands\label{inst:leiden}\and
Max-Planck-institute f{\"u}r extraterrestrische Physik, Giessenbachstra{\ss}e, D-85748 Garching, Germany\label{inst:mpe}}
                       

\abstract{How protoplanetary disks evolve is still an unsolved problem where different processes may be involved. Depending on the process, the disk gas surface density distribution $\Sigma_{\rm gas}$ may be very different and this could have diverse implications for planet formation. Together with the total disk mass, it is key to constrain $\Sigma_{\rm gas}$ as function of disk radius $R$ from observational measurements. } 
{In this work we investigate whether spatially resolved observations of rarer CO isotopologues, such as $^{13}$CO, may be good tracers of the gas surface density distribution in disks.} 
{Physical-chemical disk models with different input $\Sigma_{\rm gas}(R)$ are run, taking into account CO freeze-out and isotope-selective photodissociation. The input disk surface density profiles are compared with the simulated $^{13}$CO intensity radial profiles to check whether and where the two follow each other.} 
{For each combination of disk parameters, there is always an intermediate region in the disk where the slope of the $^{13}$CO radial emission profile and  $\Sigma_{\rm gas}(R)$ coincide. In the inner part of the disk the line radial profile underestimates $\Sigma_{\rm gas}$, as $^{13}$CO emission becomes optically thick. The same happens at large radii where the column densities become too low and $^{13}$CO is not able to efficiently self-shield. Moreover, the disk becomes too cold and a considerable fraction of $^{13}$CO is frozen out, thus it does not contribute to the line emission. If the gas surface density profile is a simple power-law of the radius, the input power-law index can be retrieved within a $\sim 20 \%$ uncertainty if one choses the proper radial range. If instead $\Sigma_{\rm gas}(R)$ follows the self-similar solution for a viscously evolving disk, retrieving the input power-law index becomes challenging, in particular for small disks. Nevertheless, it is found that the power-law index $\gamma$ can be in any case reliably fitted at a given line intensity contour around 6 K km s$^{-1}$, and this produces a practical method to constrain the slope of $\Sigma_{\rm gas}(R)$. Application of such a method is shown in the case study of the TW Hya disk.} 
{Spatially resolved $^{13}$CO line radial profiles are promising to probe the disk surface density distribution, as they directly trace $\Sigma_{\rm gas}(R)$ profile at radii well resolvable by ALMA. There, chemical processes like freeze-out and isotope selective
photodissociation do not affect the emission, and, assuming that the
volatile carbon does not change with radius, no chemical model is
needed when interpreting the observations.} 


\maketitle

\section{Introduction}
Protoplanetary disk evolution adds to the construction of  the host young star and simultaneously provides the material for the formation of planets and large rocky bodies. Mass accretion onto the central star and planet formation are therefore connected processes where the gaseous and solid disk components play important and complementary roles \citep[see][for a review]{Armitage15}. It is generally assumed that gas is the main disk component accounting for 99\% of its mass, while only 1\% is in the form of dust. Furthermore, the gaseous component is also what drives disk evolution and thus the disk structure at different stages. An open question is what the relative importance is of different mechanisms transporting the material onto the star and dispersing the disk within few Myr. Disk evolution leads to mass accretion onto the central star \citep{Hartmann16}, but part of the disk content can be dissipated by high-energy radiation driven winds or magnetic torques \citep{Alexander14,Gorti16,Bai16} and external processes such as encounters \citep[e.g.][]{Clarke93,Pfalzner05} and external photoevaporation \citep[e.g.][]{Clarke07,Anderson13,Facchini16}. Depending on which process is dominant, the disk structure will be considerably different. In particular this will impact the distribution of the gas as a function of the distance from the star, which is described by the surface density distribution, $\Sigma_{\rm gas}$. 
\cite{Morbidelli16} discuss the difference in the resulting disk structures between disk wind models and the standard viscously evolving $\alpha$-disk model. \cite{Suzuki16} find that the inner disk can be heavily depleted as winds efficiently remove angular momentum from the disk and cause large accretion onto the star. On the other hand, a different disk wind model by \cite{Bai16} predicts a surface density profile that scales inversely with the radius in the innermost 10 au, similarly to what is expected in the $\alpha$-disk case.
Reliable observational measurements of $\Sigma_{\rm gas}$ as function of radius would therefore be key to understand disk evolution and the relative importance of different processes, as well as how planet formation occurs \citep{Morbidelli16}. 

Successful attempts have been made to constrain the disk dust surface density $\Sigma_{\rm dust}$ from high S/N spatially resolved mm-continuum observations. The thermal emission of mm-sized grains is largely optically thin in protoplanetary disks and allows to directly trace the column density of the bulk of the dust \citep[e.g.][]{Andrews15,Tazzari17}. Dust and gas surface densities, however, are not necessarily the same. 
Even though they are strongly coupled, different mechanisms are
thought to affect their evolution, such as radial drift of
millimeter-sized grains modifying the dust surface density on short
timescales. This is the case e.g., in HD 163296 where the carbon monoxide (CO) emission
extends much further than the continuum dust emission \citep{deGregorio-Monsalvo13,Isella16}, although the much higher optical depth of the CO lines versus millimeter continuum also contributes significantly to this difference in appearance \citep{Dutrey98,Facchini17}.  Another example comes from the Herbig disk HD 169142, where ALMA observations show that the dust appears to be concentrated in two rings between 20--35 au and 56--83 au, whereas gas seen in CO isotopologues is still present inside the dust cavity and gap \citep{Fedele17}.  More extreme examples are transitional disks with large cavities severely depleted in dust content, but less in gas, with gas cavities smaller than dust cavities \citep{vanderMarel13,vanderMarel16a,Bruderer14,Zhang14}. Taken together, it is clear that gas and millimeter dust emission do not follow each other and that a more direct tracer of the gas surface density profile is needed.

The main gaseous component is molecular hydrogen (H$_2$), whose emission lines at near- and mid-infrared wavelengths are very weak and are superposed on strong continuum emission \citep[e.g.][]{Thi01,Pascucci06}. In contrast, CO pure rotational transitions at millimeter wavelengths are readily detected \citep[e.g.][]{Dutrey96,Thi01,Dent05,Panic08}.
CO is the second-most abundant molecule, with a chemistry that is very well studied and implemented in physical-chemical codes. Owing to their optically thin emission lines, less abundant CO isotopologues are considered better tracers of the gas column than $^{12}$CO \citep{Miotello14,Miotello16,Schwarz16}. This paper investigates whether spatially resolved observations of such lines may be good probes of the gas surface density distribution, once freeze-out and isotope selective photodissociation are taken into account.

Thanks to the unique sensitivity of ALMA, large surveys of disks in different star-forming regions are being carried out to trace gas and dust simultaneously \citep[][]{Ansdell16, Barenfeld16, Eisner16, Pascucci16}. In particular, CO isotopologues pure rotational lines have been targeted but lines are weaker than expected and at the short integration time of typically one minute the line data have low S/N. This may be interpreted as lack of gas due to fast disk dispersal, or as lack of volatile carbon that leads to faint CO lines \citep{Favre13,Ansdell16,Schwarz16,Miotello17}. As a consequence, $^{13}$CO lines are optically thin towards a larger part of the disk and become much better tracers of the gas columns than C$^{18}$O lines, which were thought to be preferred pre-ALMA, but are often undetected. Furthermore, $^{13}$CO is less affected by isotope selective photodissociation than C$^{18}$O \citep{Miotello14}. The focus of this paper is thus on $^{13}$CO, rather than on other isotopologues. Recently, \cite{Zhang17} have proposed the optically thin lines of $^{13}$C$^{18}$O, successfully detected in the TW Hya disk, to be a good probe of $\Sigma_{\rm gas}(R)$ at small radii. We briefly investigate this possibility from our modeling perspective using another low abundance isotopologue C$^{17}$O.

\cite{Williams16} have presented a method to extract disk gas surface density profiles from $^{13}$CO ALMA observations. Assuming a self-similar density distribution as given by viscous evolution theory, the authors create a large gallery of parametrized models, which they use to extract the disk parameters with a MCMC analysis. Although we share the same scientific question, this paper follows a different approach, i.e., we fit a power-law profile to our simulated $^{13}$CO radial profiles as if they were observations, in order to recover the power-law index $\gamma$. Our analysis shows that it may be difficult to infer both $\gamma$  and the critical radius independently for small disks.

In Section \ref{modeling}, the physical-chemical modeling is presented. The results of the modeling investigation are presented in Section \ref{results} and their implications are discussed in Section \ref{discussion}.

\section{Modeling}
\label{modeling}
\subsection{DALI}

The physical-chemical code DALI is used for the disk modeling \citep[Dust And LInes,][]{Bruderer12,Bruderer13}. 
DALI calculates the dust temperature, $T_{\rm dust}$, and local continuum radiation field from UV to mm wavelengths with a 2D Monte Carlo method, given an input disk physical structure and a stellar spectrum as source of irradiation. Then, a time-dependent  (1 Myr) chemical network simulation is run. 
Subsequently, the gas temperature, $T_{\rm gas}$, is calculated through an iterative balance between heating and cooling processes until a self-consistent solution is found and the non-LTE excitation of the molecules is computed. In this paper, the final outputs are spectral image cubes of CO isotopologues computed by a ray-tracer module. Throughout this paper, the $J$=3-2 line is produced and analyzed,  but results are similar for $J$=2-1. Line intensities are provided in K km s$^{-1}$, rather than Jy beam$^{-1}$ km s$^{-1}$, in order to not depend on the beam size.  The line intensity can be converted to Jy beam$^{-1}$ km s$^{-1}$ using the relation in Appendix B of \cite{Bruderer13}. For the ray-tracing, a distance of 150 pc is assumed.

DALI has been extensively tested and benchmarked against observations \citep[see e.g.,][]{Fedele13} and a detailed description of the code is presented by \cite{Bruderer12} and \cite{Bruderer13}. For this work, a complete treatment of CO freeze-out and isotope-selective processes is included \citep[for more detail, see][]{Miotello14,Miotello16}. The volatile carbon abundance is assumed to be high, [C]/[H]=$1.35\times10^{-4}$, and constant across the disk.

\begin{figure*}
   \resizebox{\hsize}{!}
             {\includegraphics[width=2\textwidth]{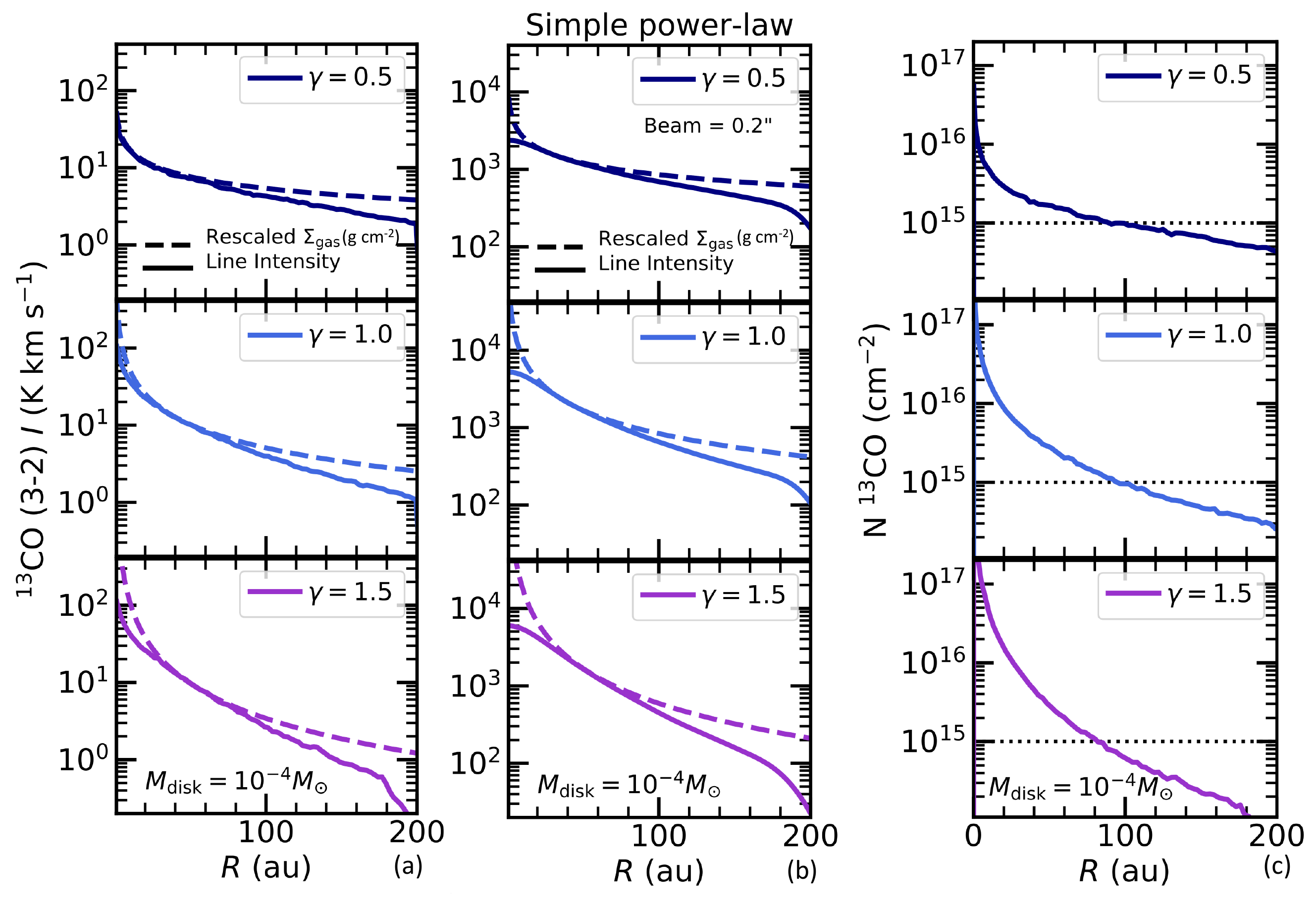}}
      \caption{\emph{Panel a -- }$^{13}$CO line intensity radial profiles (solid lines) obtained with three representative disk models with input surface density distribution $\Sigma_{\rm gas}$ (dashed lines) chosen to be a simple power-law (see Eq. \ref{sigma_eq_2}). The model parameters are $R_{\rm out}=200$ au, $M_{\rm disk}=10^{-4}M_{\odot}$ and $\gamma=0.8 ,1,1.5$ shown in dark blue, light blue, and purple respectively (top, middle, and bottom panels). No beam convolution is applied to the simulated images and a distance of 150 pc is assumed. \emph{Panel b -- } $^{13}$CO line intensity radial profiles compared with input surface density profiles as in panel a, but convolved with a 0.2" beam.
\emph{Panel c -- }Column densities of gas-phase $^{13}$CO calculated from the surface to the midplane shown as function of the disk radius for the three representative models with simple power-law surface density. The dotted black line indicates the column density at which CO self-shielding becomes inefficient ($N = 10^{15} \rm cm^{-2}$). Ice-phace CO column densities are not shown in the plot as they are below the values shown her.}
       \label{profiles_pwrl}
\end{figure*}

\subsection{Grid of models}
\label{models}

The disk surface density distribution is often parametrized by an exponentially tapered power-law function, following the prescription proposed by 
\cite{Andrews11}. Physically this represents a viscously evolving disk, where the viscosity is expressed by $\nu \propto R^{\gamma}$ \citep{Lynden-BellPringle74,Hartmann98}. The self-similar surface profile is expressed by:
\begin{equation}
\Sigma_{\rm gas}=\Sigma_c \,\left( \frac{R}{R_c} \right) ^{-\gamma} \exp \, \left[ - \,\left( \frac{R}{R_c} \right) ^{2-\gamma} \right] \ ,
\label{sigma_eq}
\end{equation}
where $R_c$ is the critical radius and $\Sigma_c$ is the surface density at the critical radius. 
This parametrization has often been employed to model disks with DALI \citep[see e.g.,][]{Miotello14,Miotello16}. Adopting an exponential taper to the power-law profile of the surface density distribution (see Eq. \ref{sigma_eq}), as suggested by viscous evolution theory, has the inconvenience that the profile slope depends on two free parameters, $R_{\rm c}$ and $\gamma$, which can be difficult to disentangle with the current low S/N of the data in the outer regions of disks. A simplification comes from the assumption that $\Sigma_{\rm gas}$ has a pure power-law dependence with radius. In this case the power-law index $\gamma$ is left as a single free parameter:
\begin{equation}
\Sigma_{\rm gas}=\begin{cases}
\Sigma_c \,\left( \frac{R}{R_c} \right) ^{-\gamma} & R \le R_{\rm out},\\
0 & R > R_{\rm out}. \end{cases}
\label{sigma_eq_2}
\end{equation}
In this work both parametrizations have been employed to design the input density structures. 

\begin{table}[tbh]
\caption{Parameters of the disk models.}
\label{Tab_simple_pwr}
\centering
\begin{tabular}{lll}
\hline\hline
Parameter	 &\multicolumn{2}{c}{Value} \\
\hline
& Self similar & Power-law\\
\hline
$\gamma$ & 0.8, 1, 1.5 & 0.5, 1, 1.5\\
$R_{\rm c}$ & 30, 60, 200 au & -- \\
$R_{\rm out}$ & 500 au & 100, 200 au \\
$M_{\rm gas}$ & $10^{-5}, 10^{-4}, $ & $10^{-5}, 10^{-4},$ \\
& $10^{-3}, 10^{-2} M_\odot$ & $10^{-3}, 10^{-2} M_\odot$\\
\hline
\end{tabular}
\end{table}

First, some of the model results presented in \cite{Miotello14} using the self similar profile have been analyzed. The free parameters are then $R_{\rm c}$, $\gamma$, and $M_{\rm disk}$, whose values are reported in column 2 of Table \ref{Tab_simple_pwr}. A value for $R_{\rm out}$ is also reported, but this is simply needed for the simulation and does not have any effect on the disk structure, as the exponential cut-off truncates the disk at smaller radii. Parameters defining the disk vertical structure are reported in Table 1 of \cite{Miotello14} and simulate a medium-flared disk with large dust grains settled. From here on, such disk models where the input surface density distribution is set by Eq. \ref{sigma_eq} will be called \emph{self-similar} disk models.

Additional models have then been run with the simple power-law surface density structure (see eq. \ref{sigma_eq_2}). In this case the free parameters are $R_{\rm out}$, $\gamma$, and $M_{\rm disk}$ as shown in column 3 of Table \ref{Tab_simple_pwr}. From here on, disk models where the input surface density distribution is set by Eq. \ref{sigma_eq_2} will be called \emph{simple power-law} disk models.

In order to investigate different disk mass regimes and then compare with recent observations of CO isotopologues in protoplanetary disks \citep{Ansdell16,Ansdell18}, a range of models from less massive to high mass disks have been run (see Tab. \ref{Tab_simple_pwr}). These observations also show moderately extended emission, therefore the outer radius in the models has been set to 200 au. The raytracing of all models presented here is carried out assuming the disk to be at a distance of 150 pc, representative of that of the nearby  star-forming regions. Furthermore, the output synthetic images have either been left unconvolved or convolved with a moderate resolution beam of 0.2$^{\prime\prime}$. In the first case, the resolution is that of the radial grid assumed for the simulation. The calculation is carried out on 75 cells in the radial direction (logarithmically spaced up to 30 au and then linearly spaced for a radial resolution of $\sim 10$ au) and 60 in the vertical direction models \citep{Miotello14}. In the new models with the simple power-law surface density structure, a denser grid with 95 cells in the radial direction (improved radial resolution in the outer disk of $\sim 3$ au) and 80 in the vertical direction was chosen. 

All models presented in this work are T Tauri-like disk models. The stellar spectrum
is assumed to be a black-body at a temperature $T_{\rm eff}=4000$ K with excess UV due to accretion and the stellar luminosity is $L_{\rm bol}=1 L_{\odot}$ \citep[see][]{Miotello14,Miotello16}. In order to explore the effect of different stellar properties on the results of this work, we have compared the self similar disk models with some of the Herbig-like disk models presented in \cite{Miotello16}, where $T_{\rm eff}=10000$ K and $L_{\rm bol}=10 L_{\odot}$. We find that stellar temperature and luminosity do not quantitatively affect the results found with the T-Tauri like disk models, except for pushing the preferred fitting region further out in the disk for massive disk models (see Sec. \ref{slope-pivot-point}). Such massive Herbig disk models ($M_{\rm disk}=10^{-3}M_{\odot}$) show radial intensity profiles which are a factor of 1.6 brighter than those found for the respective T-Tauri models. 

\begin{figure*}
   \resizebox{\hsize}{!}
             {\includegraphics[width=1.\textwidth]{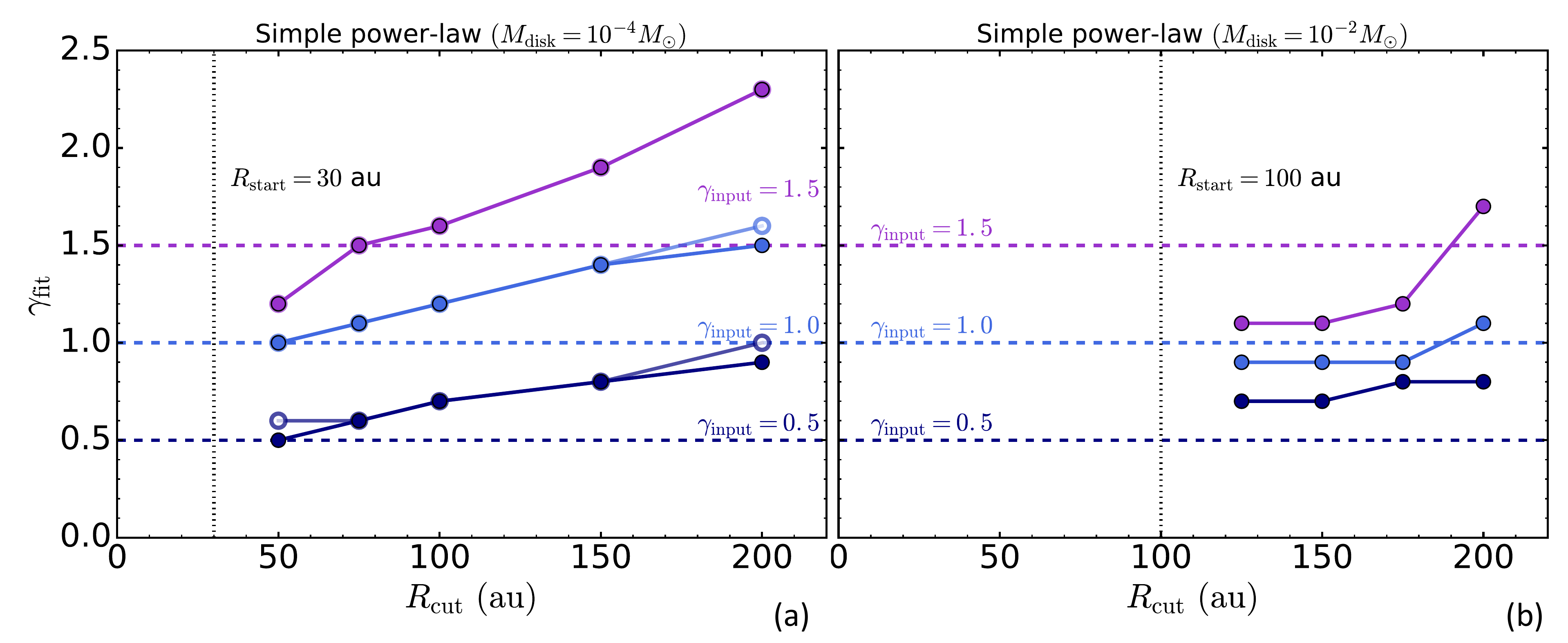}}
      \caption{Results of the power-law fitting of the simulated $^{13}$CO line intensity profiles obtained with the simple power-law models ($M_{\rm disk}=10^{-4}M_{\odot}$ and $M_{\rm disk}=10^{-2}M_{\odot}$ in panel a and b respectively). The values of the fitted power-law index $\gamma_{\rm fit}$ are shown by the filled dots as a function of the radial range over which the fitting was carried out for the high resolution images (0.01" beam). The empty symbols in panel (a) show the fitted power-law index $\gamma_{\rm fit}$ when the simulated images are convolved with a 0.2" beam. The model input power-law indexes are shown by the dashed lines. The dotted line shows the starting point of the fitting $R_{\rm start}$. }
       \label{gamma_fit_pwrl}
\end{figure*}

\section{Results}
\label{results}

The first test is to compare the input disk surface density profiles for the DALI models with the simulated $^{13}$CO profiles, as if they were observed at infinite resolution, i.e. not convolved with any beam. In practice, the image resolution is the physical grid resolution in our models (see Sec. \ref{models}). The next step is then to convolve the simulated $^{13}$CO profiles with a typical ALMA observation beam. The results obtained in the two cases, where $\Sigma_{\rm gas}$ is parametrized by a simple power-law or following the self-similar solution are presented below.

 \subsection{Simple power-law}
\label{sec_pwrl}

The input disk surface density profiles, assumed to be a power-law function of the radius (see Eq. 2), are compared with the simulated $^{13}$CO profiles for three representative models ($M_{\rm disk}=10^{-4} M_{\odot}$, $R_{\rm out}=200$ au, $\gamma=0.8,1,1.5$) and are shown in panel (a) of Fig. \ref{profiles_pwrl}. The surface density profiles (in g cm$^{-2}$) have been rescaled by a constant factor in order to visually match the line intensity profiles. The disk can be divided in three regions: an inner part in which the $^{13}$CO radial profile underestimates $\Sigma_{\rm gas}$, a central zone where the two coincide, and an outer region where the line emission drops and deviates from the surface density distribution (see case with $\gamma=1.5$, for clarity). At small radii the divergence is caused by the fact that $^{13}$CO emission becomes optically thick as the columns are very high. At large radii the $^{13}$CO column densities become too low, lower than $10^{15}$ cm$^{-2}$, and $^{13}$CO is not able to efficiently self-shield \citep{VDB88,Visser09}. 

There are two ways to further investigate these cases. The column densities of gas-phase $^{13}$CO calculated from the surface to the midplane are shown as function of the disk radius in panel (c) of Fig. \ref{profiles_pwrl} in blue, light blue and purple for models with $\gamma=0.8 ,1,1.5$ respectively. The column density can be compared with the line luminosity profiles presented in panel (a) of Fig. \ref{profiles_pwrl}. The radius where the line luminosity profiles decrease and deviate from the surface density distribution is similar to that at which the $^{13}$CO column densities drop below $10^{15}$ cm$^{-2}$. This illustrates that inefficient CO self-shielding affects CO isotopologues intensity profiles and needs to be considered in the disk outer regions.

As reported in Table \ref{Tab_simple_pwr}, less and more massive models have been run for the simple power-law case. We observe similar trends, but the radial location of the three regions described above are radially shifted to smaller or larger radii, depending on disk mass (see Fig. \ref{prof_masses}). Compared with the representative model with mass $M_{\rm disk}=10^{-4} M_{\odot}$, for $M_{\rm disk}=10^{-5} M_{\odot}$ the simulated radial intensity profiles better follow $\Sigma_{\rm gas}$ at smaller radii as the line is less optically thick. On the other hand $^{13}$CO column densities smaller than $N=10^{15}\rm cm^{-2}$ are reached at smaller radii. The opposite happens for $M_{\rm disk}=10^{-2} M_{\odot}$, where the line emission is optically thick up to almost 100 au, but CO self-shielding becomes inefficient much further out, around 200 au.

\begin{figure*}
\centering
   \resizebox{0.7\hsize}{!}
            { \includegraphics[width=0.7\textwidth]{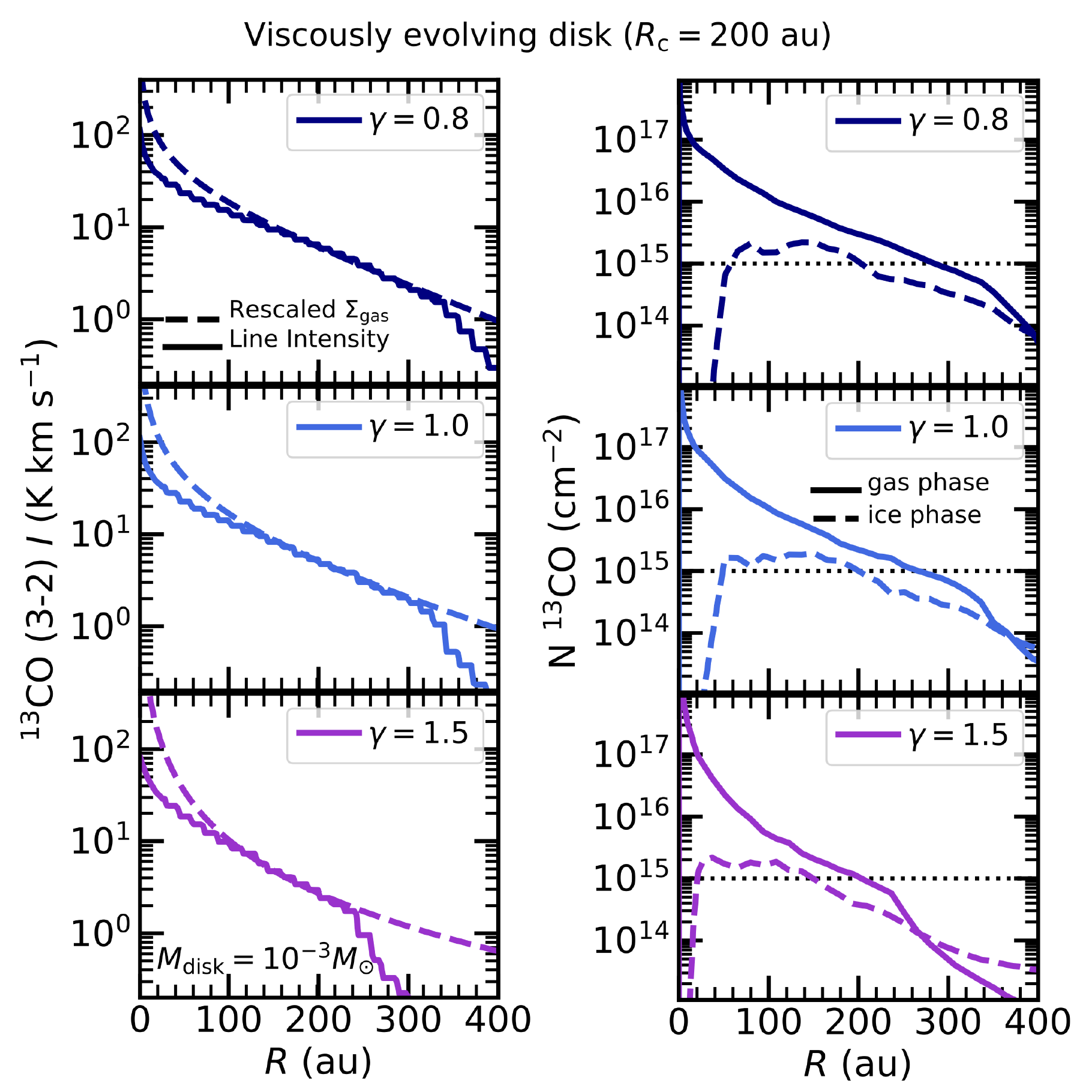}}
      \caption{\emph{Panel (a) -- }$^{13}$CO line intensity radial profiles (solid lines) obtained with three representative disk models with input surface density distribution $\Sigma_{\rm gas}$ (dashed lines) given by the viscously evolving disk model (see Eq. \ref{sigma_eq}). The model parameters are $R_{\rm c}=200$ au, $M_{\rm disk}=10^{-3}M_{\odot}$ and $\gamma=0.8 ,1,1.5$ shown in dark blue, light blue, and purple respectively (top, middle, and bottom panels). \emph{Panel (b) -- }Column densities of gas-phase (solid lines) and ice-phase (dashed lines) $^{13}$CO calculated from the surface to the midplane shown as function of the disk radius for the three representative self-similar disk models. The dotted black line indicates the column density at which CO self-shielding becomes inefficient ($N = 10^{15} \rm cm^{-2}$).} 
       \label{profiles}
\end{figure*}

 \subsubsection{Fitting of the surface density power-law index $\gamma$}
\label{fit}

We now address the question of whether it is possible to retrieve the underlying surface density power-law index $\gamma$ by fitting the line intensity profiles. We refer to the power-law index used as input in DALI as $\gamma_{\rm input}$ and we label the fitted value  $\gamma_{\rm fit}$.

The fit of a power-law profile to the simulated intensity profiles is performed as a linear fit in the $\log{I}$-$\log{R}$ space. As it is clear from Fig. \ref{profiles_pwrl} that $^{13}$CO radial intensity profiles follow $\Sigma_{\rm gas}(R)$ only in a particular region, the fit is not carried out over the whole extent of the disk, but over a radial range that spans from $R_{\rm start}$ to $R_{\rm cut}$. The starting point $R_{\rm start}$ is kept fixed and chosen to be just outside the inner region where the emission is optically thin, typically 30--40 au. The range of radii over which the fitting is performed is then varied by changing the value of $R_{\rm cut}$.

The results of the fitting are presented in Fig. \ref{gamma_fit_pwrl} for the single power-law case. The dashed lines illustrate the value of $\gamma_{\rm input}$, while the colored dots show the value of $\gamma_{\rm fit}$ if different $R_{\rm cut}$ are chosen for the fitting. The starting radius for the fitting procedure $R_{\rm start}$ is indicated by the dotted vertical line.  

For low mass disk models ($M_{\rm disk}=10^{-4}M_{\odot}$, panel a of Fig. \ref{gamma_fit_pwrl}), the power-law index $\gamma_{\rm input}$ can be retrieved within 20$\%$ of uncertainty in a range of radii that goes from 30 -- 40 au to $\sim 100$ au, the typical scales probed with ALMA. Furthermore, the retrieved $\gamma_{\rm fit}$ usually overestimates the input value, except for the $\gamma=1.5$ case. Fixing $R_{\rm start}$ to either 30 au or 40 au  does not change qualitatively the outcome of the fit. For $R_{\rm cut}$ larger than 100 au, the fitted values deviate significantly from $\gamma_{\rm input}$ as one enters the region where self-shielding becomes inefficient and the intensity profiles deviate from $\Sigma_{\rm gas}$. For higher mass disk models ($M_{\rm disk}=10^{-2}M_{\odot}$, panel b of Fig. \ref{gamma_fit_pwrl}) one needs to chose a larger $R_{\rm start}$ of 100 au, as $^{13}$CO is optically thick much further out than for the low mass disk case. Furthermore, the uncertainties on $\gamma_{\rm fit}$ are larger.

In order to choose the correct radial range over which to perform the fit, one would need to know in which mass regime the observed disk is. More precisely, one would need to constrain the total $^{13}$CO gas mass. For faint observed $^{13}$CO fluxes, \cite{Miotello16} have found a linear relation between line emission and total disk mass. This could be used to constrain $M_{\rm disk}$. For brighter observed $^{13}$CO fluxes, this relation is less reliable and one could employ the dust masses obtained from continuum emission, multiplied by the canonical factor of 100. This would provide an upper limit to the disk mass (see Sec. \ref{discussion} for discussion on carbon depletion). Another possibility is to combine $^{13}$CO and C$^{18}$O line fluxes \citep{Miotello16}, if the latter are available from the same observation.

\begin{figure*}
\centering
             \includegraphics[width=1.5\columnwidth]{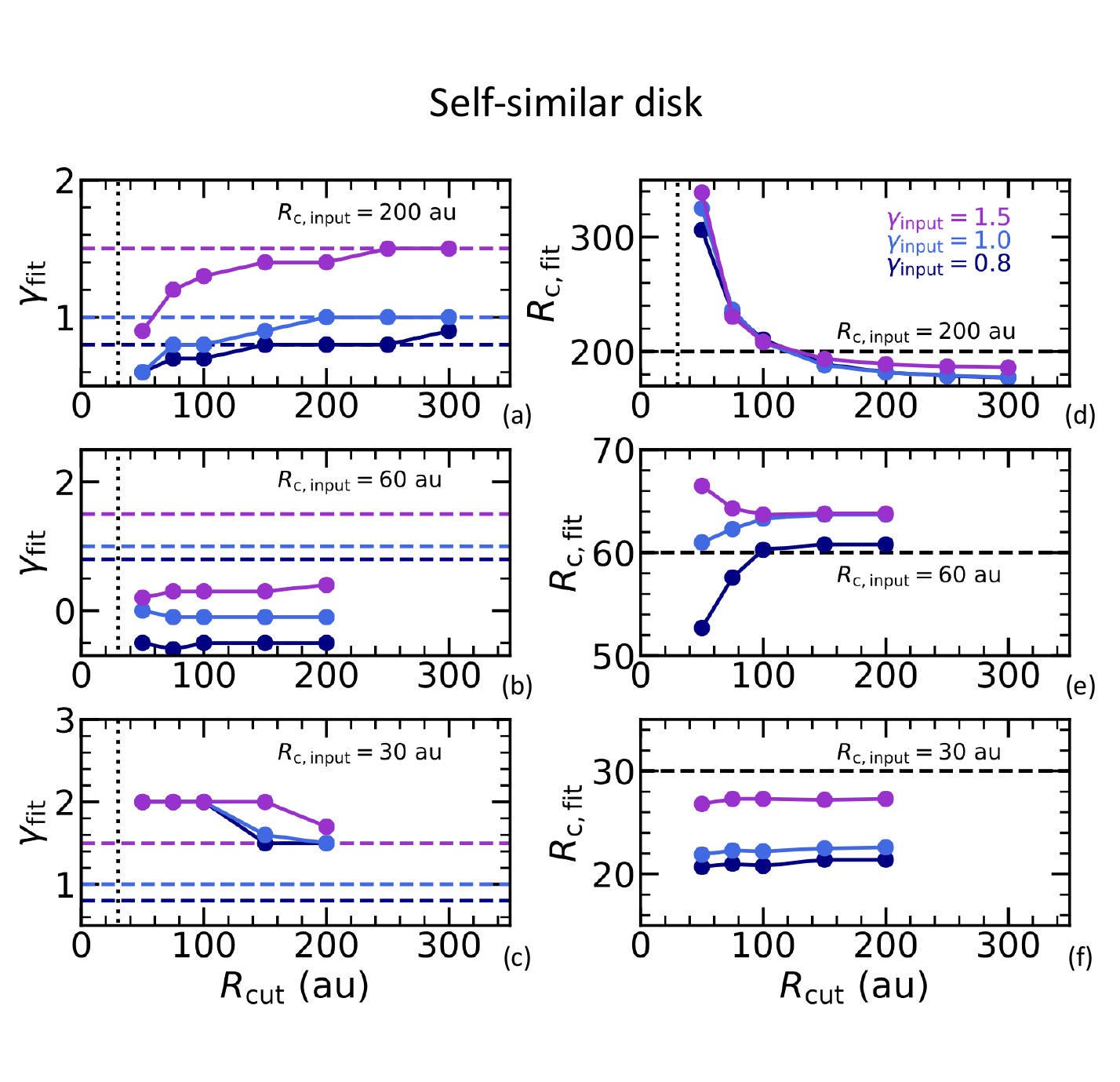}
      \caption{Results of the power-law fitting of the simulated $^{13}$CO line intensity profiles obtained with the self-similar disk models with $M_{\rm disk}=10^{-4} M_{\odot}$. The values of the fitted power-law index $\gamma_{\rm fit}$ are shown by the filled dots as a function of the radial range in which the fitting was carried out. The model input power-law indexes are shown by the dashed lines. The dotted line shows the starting point of the fitting $R_{\rm start}$.}
       \label{gamma_fit_exp}
\end{figure*}

\subsubsection{Convolution with a beam}

Finally, the simulated line intensity radial profiles need to be convolved with a synthesized beam to check how the trends are affected by this operation. Observations taken from recent ALMA disk surveys have moderate angular resolution, between 0.2" and 0.3", corresponding to 30 -- 40 au diameter at 150 pc \citep[e.g.]{Ansdell16}. To mimic such observations we convolve our simulated profiles with a 0.2" beam. The convolved $^{13}$CO line intensity radial profiles are shown in panel b of Fig.  \ref{profiles_pwrl}. 

As in Sec. \ref{fit}, a power-law fitting of the convolved radial intensity profiles has been carried out. The starting point $R_{\rm start} $ has been fixed to 30 au and the results are shown in panel (a) of Fig. \ref{gamma_fit_pwrl}. The values of $\gamma_{\rm fit}$ do not qualitatively differ from those found with the unconvolved radial profiles, shown in panel (a) of Fig. \ref{gamma_fit_pwrl}. Even with a beam convolution of 0.2" the power-law index can be retrieved to an accuracy of $\sim 20 \%$, if the fit is carried out in a region that goes from $\sim$30 au to $\sim$100 au.

\subsection{Self-similar disk models}
\label{sec:exp}

Similar trends as those presented in Sec. \ref{sec_pwrl} have been found for the self-similar case.
The input disk surface density profiles compared with the simulated $^{13}$CO profiles for three representative self-similar models ($M_{\rm disk}=10^{-3} M_{\odot}$, $R_{\rm c}=200$ au, $\gamma=0.8,1,1.5$) are presented in panel (a) of Fig. \ref{profiles}. The surface density profiles have been rescaled by a constant factor in the plots to visually match the line intensity profiles. 
As for the simple power-law case it is possible to identify three regions: the inner disk in which $^{13}$CO lines are optically thick and the emission profile is flatter than the $\Sigma_{\rm gas}$ profile, a central region in which the two profiles match, and the outer disk where the line intensity underestimates the surface density distribution

In the self-similar models a secondary effect adds to inefficient CO self-shielding to reduce the emission in the outer disk. As shown in Fig. \ref{n13CO}, the simple power-law representative model (panel b) for a less massive disk ($M_{\rm disk} = 10^{-4}M_{\odot}$) is warmer than the self-similar  representative disk model (panel a, $M_{\rm disk} = 10^{-3}M_{\odot}$). In this second case, the dust temperature $T_{\rm dust}$ drops below 20 K in the midplane at radii larger than 50 au and a significant fraction of $^{13}$CO is frozen out, thus not contributing to the line emission (see panel a of Fig. \ref{n13CO}). Moreover, the inner region where the emission lines are optically thick is smaller compared to the power-law case, and the radial range over which the line intensity profile follows $\Sigma_{\rm gas}$ is shifted toward smaller radii. 

The column density, reported in panel b of Fig. \ref{profiles}, can be compared with the line intensity profiles presented in panel a. The radius where the line intensity profiles decrease and deviate from the surface density distribution is similar to that at which $^{13}$CO column densities drop below $10^{15}$ cm$^{-2}$ (at $\sim$300 au for $\gamma=0.8$, $\sim$250 au for $\gamma=1.0$, and $\sim$200 au for $\gamma=1.5$).

Similarly to the simple power-law case, simulated $^{13}$CO radial intensity profiles have been fitted with an exponentially tapered power-law (see Equation \ref{sigma_eq}) to try and retrieve the power-law index $\gamma_{\rm input}$ and the critical radius $R_{\rm c,input}$. This procedure is applied to models where $M_{\rm disk}=10^{-4}M_{\odot}$, in order to directly compare with the results found in the simple power-law case, instead of $M_{\rm disk}=10^{-3}M_{\odot}$, as discussed above.  The intensity profiles for the lower mass ($M_{\rm disk}=10^{-4}M_{\odot}$) self-similar disk models are reported in Fig. \ref{exp_low_mass} in Appendix \ref{lower_mass}.
Fit parameters are much more ambiguous when fitting the intensity profiles obtained with the self-similar disk models. As shown in Fig.  \ref{gamma_fit_exp}, the retrieved $\gamma_{\rm fit}$ and $R_{\rm c,fit}$ can significantly deviate from the original $\gamma_{\rm input}$. If $R_{\rm c,input}$ is 200 au, it is still possible to distinguish $\gamma_{\rm fit}$ between models with different $\gamma_{\rm input}$ (panel a of Fig. \ref{gamma_fit_exp}). This holds in particular if $R_{\rm cut}> 100$ au, in which case $\gamma_{\rm input}$ is retrieved within 20\% and $R_{\rm c,input}$ within 12\%, consistent with \cite{Williams16} large HD163296 disk. If instead $R_{\rm c}$ is smaller, i.e., 30 au or 60 au (panel b and c of Fig. \ref{gamma_fit_exp} respectively), this is not true as $\gamma_{\rm input}$ cannot be reliably retrieved. It is still possible to obtain a good estimate for $R_{\rm c}$, which can be retrieved within 13\% and 30\% if $R_{\rm c,input}=60, 30$  au respectively. The complication is given by the fact that for smaller values of $R_{\rm c}$ the line intensity radial profile follows the $\Sigma_{\rm gas}$ profile over a very limited radial range and the fitting does not perform well enough. The effects of either optical thickness or inefficient self-shielding dominate throughout most of the disk's extent and the radial line intensity profile does not resemble an exponentially-tapered power-law.
In summary, the obtained $\gamma_{\rm fit}$ has a poor relation with $\gamma_{\rm input}$ if $R_{\rm c,input}$ is small (i.e., 30, 60 au). On the other end, it is always possible to retrieve $R_{\rm c}$ with moderate uncertainty.

 Recent ALMA disk surveys \citep[e.g.][]{Ansdell16,Pascucci16} have shown that typical protoplanetary disks are usually fainter and less extended than previously thought. Therefore, the typical  sensitivity of these observations does not allow to detect the regions where the exponential taper would dominate, if $\Sigma_{\rm gas}$ were described by Eq. 1. Moreover, our results show that in such external regions the inefficiency of CO self-shielding prevents CO isotopologues to be used to constrain the surface density distribution. Therefore, we consider the power-law prescription as a simplification of the self-similar disk model. Both behave in the same way, i.e. as a simple power-law disk model, in the region where CO isotopologue can be reliably employed to trace $\Sigma_{\rm gas}$.
 
\subsection{Inner disk surface density profile from C$^{17}$O line intensity radial profiles}
Our models show that $^{13}$CO is not a good tracer of $\Sigma_{\rm gas}$ in the inner 30--40 au.
Recently, \cite{Zhang17} have used the rarer $^{13}$C$^{18}$O isotopologue to try determining the surface density profile in the inner regions of the well studied TW Hya disk, inside the CO snowline. Another rather low abundance CO isotopologue is C$^{17}$O, expected to be around 24 times more abundant than $^{13}$C$^{18}$O based on isotope ratios. We thus analyze the intensity profiles obtained for C$^{17}$O to check how they can be used to infer the surface density profile in the disk inner regions.

\begin{figure*}
   \resizebox{\hsize}{!}
             {\includegraphics[width=2\textwidth]{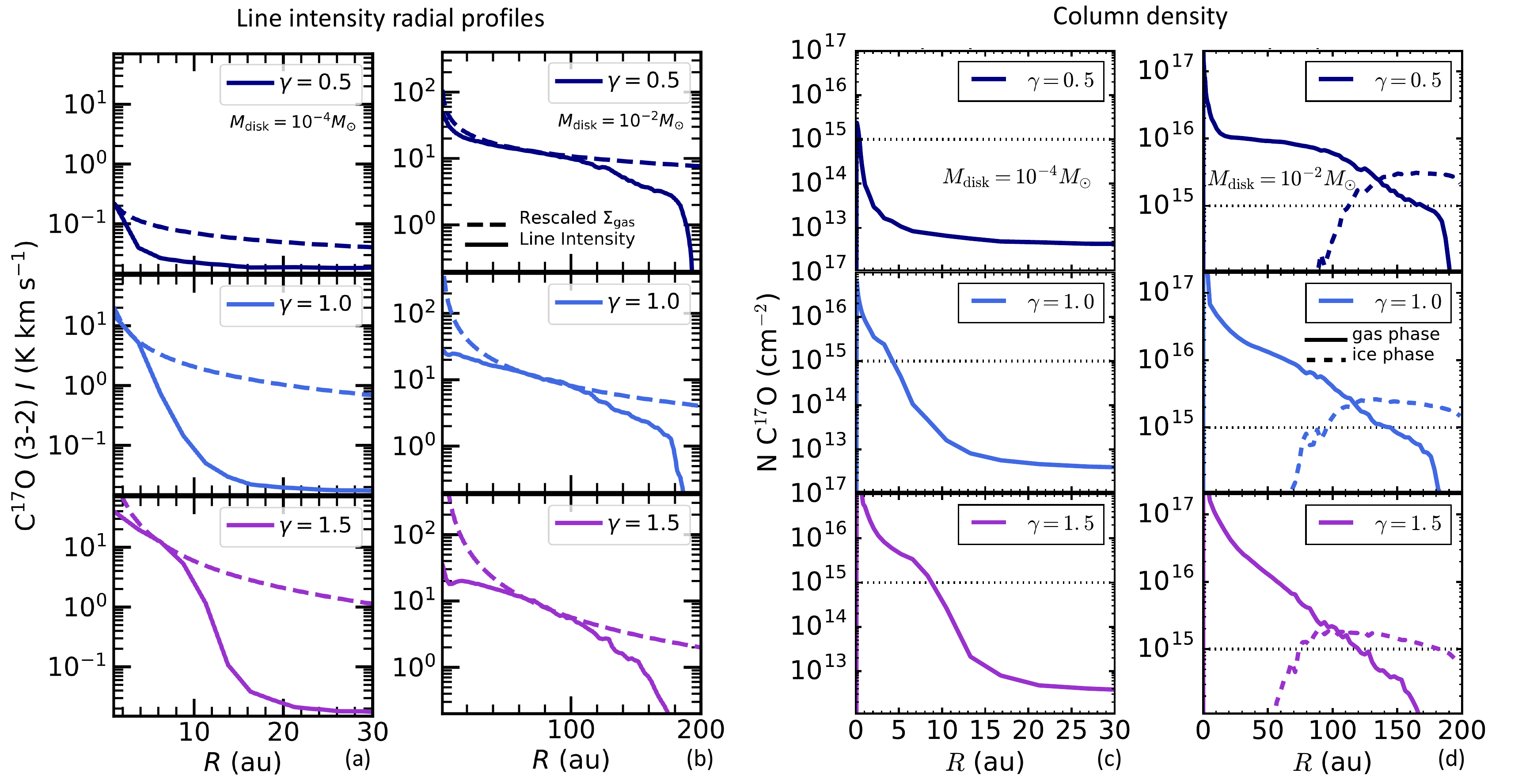}}
      \caption{\emph{Panels (a) and (b) -- }C$^{17}$O line intensity radial profiles (solid lines) obtained with six disk models with input surface density distribution $\Sigma_{\rm gas}$ (dashed lines) chosen to be a simple power-law (see Eq. \ref{sigma_eq_2}). The model parameters are $R_{\rm out}=200$ au, $M_{\rm disk}=10^{-4} M_{\odot}$ (left),$\, 10^{-2} M_{\odot}$ (right) and $\gamma=0.8 ,1,1.5$ shown in dark blue, light blue, and purple respectively (top, middle, and bottom panels). \emph{Panels (c) and (d) -- }Column densities of gas-phase (solid lines) and ice-phase (dashed lines) of C$^{17}$O calculated from the surface to the midplane shown as function of the disk radius for six models with simple power-law surface density ($R_{\rm out}=200$ au, $M_{\rm disk}=10^{-4},\, 10^{-2} M_{\odot}$). Top, middle, and bottom panels represent the models with $\gamma=0.8,1,1.5$ respectively. The dotted black lines indicate the column density at which CO self-shielding becomes inefficient ($N = 10^{15} \rm cm^{-2}$).}
       \label{profiles_C17O}
\end{figure*}

The simulated C$^{17}$O intensity profiles and column density profiles for six power-law disk models ($R_{\rm out}=200$ au, $M_{\rm disk}=10^{-4},\, 10^{-2} M_{\odot}$ and $\gamma=0.8 ,1,1.5$) are presented in Fig. \ref{profiles_C17O}. For the $10^{-4} M_{\odot}$ mass disk models, the intensity profiles follow the shape of $\Sigma_{\rm gas}$ in the very inner disk ($R<10$ au, see panel a of Fig. \ref{profiles_C17O}), apart for the steeper $\gamma=1.5$ model where optical thickness starts to play a role for $R<5$ au. On the other hand C$^{17}$O column densities drop below $10^{15}\rm cm^{-2}$ very early, at $R<10$ au. Thus, any information about $\Sigma_{\rm gas}$ from C$^{17}$O intensity profiles is lost for the outer disk regions.

The picture changes if the results of the more massive ($M_{\rm disk}=10^{-2} M_{\odot}$) disk model  are analyzed. C$^{17}$O emission is optically thick out to 50 au, except for the model with $\gamma=0.5$, where $\Sigma_{\rm gas}$ is less steep in the inner disk (see panel b in Fig. \ref{profiles_C17O}). Therefore C$^{17}$O is not a good tracer of the surface density distribution in the inner disk in massive disks. An even rarer CO isotopologue, as $^{13}$C$^{18}$O, should be used in these cases, as was done for TW Hya \citep{Zhang17}. A possible complication comes from the fact that its faint emission may be shielded by optically thick continuum emission, and this would prevent  $^{13}$C$^{18}$O from being a good tracer of the gas column density. 

As for $^{13}$CO, it is possible to fit the simulated intensity profiles with a power-law and check for which conditions $\gamma_{\rm fit}$ resembles $\gamma_{\rm input}$ well enough.

\subsection{A simple prescription to fit $\gamma$: the "slope-pivot-region}
\label{slope-pivot-point}
\begin{figure*}
   \resizebox{\hsize}{!}
             {\includegraphics[width=1.\textwidth]{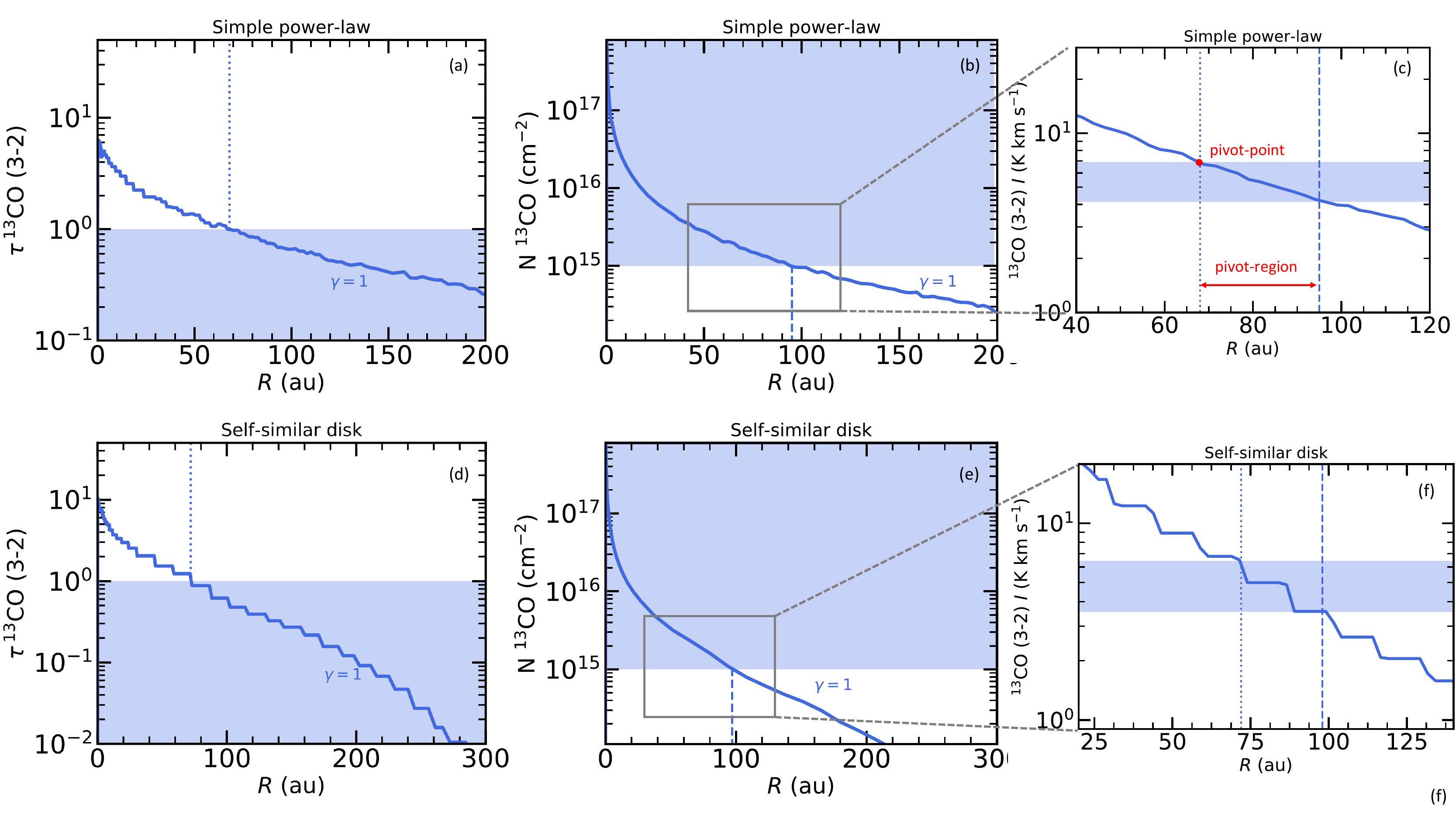}}
      \caption{Total $^{13}$CO line optical depth $\tau$ (panels a, and d), $^{13}$CO column density (panels b, and e), and $^{13}$CO intensity radial profiles for simple power-law disk models (top panels) and self-similar disk models (bottom panels). The total disk mass is set to $M_{\rm disk}=10^{-4} M_{\odot}$ and $\gamma=1$. The shaded regions in panels (a) and (d) show the range where $\tau<1$, which holds for radii larger than those shown by the dotted lines. The shaded region in panels (b) and (e) shows the region where $^{13}$CO column densities are larger than $10^{15}$ cm$^{-2}$, i.e. for radii smaller than those shown by the dashed lines. The shaded regions in panels (c) and (f) show the range of $^{13}$CO line intensity for which the conditions $\tau<1$ and $N^{13}$CO $>10^{15}$ cm$^{-2}$ hold. The fit of $\gamma$ is performed over the "slope-pivot-region" between the dotted and dashed lines, shown by the red arrow in panel (c).}
       \label{pivot_point}
\end{figure*}

The fitting method of the surface density power-law index presented in Sec. \ref{sec_pwrl} is not straightforward to be applied to observations, as this approach involves the choice of two parameters, $R_{\rm start} $ and $R_{\rm cut}$, which wold be difficult to determine.
On the other hand, there is a relation between the "slope-pivot-point" (see Fig. \ref{pivot_point}), where $^{13}$CO radial intensity profile starts to deviate from the surface density profile due to optical depth, and the column density. The slope-pivot-point will always be close to where $\tau=1$, thus around the same intensity modulo temperature effects, which however are not significant for low-$J$ CO lines. One can then perform the fit of $\gamma$ between the slope-pivot-point and the point where the $^{13}$CO column density is larger than $10^{15}$ cm$^{-2}$. Such radial region will always be around the same narrow range of line intensity.\\

More specifically, at the radius where $\tau_{^{13}\rm CO} = 1$, the $^{13}\rm CO$ intensity is roughly 6 K km s$^{-1}$ (see Fig. \ref{pivot_point}).
This results from the line intensity, which is controlled by a combination of the line opacity and 
the temperature. The intensity at the frequency $\nu$ of a gas column with line opacity $\tau_{\nu}$ and uniform temperature $T$ is
\begin{equation}
I_{\nu} = B_{\nu}(T) \cdot (1-e^{-\tau_{\nu}}) 
\end{equation}
where $B_{\nu}(T)$ is the Planck function. Thus, for $\tau_{\nu} = 1$, $I_{\nu}$ depends only on the temperature. In the Rayleigh-Jeans approximation, $I_{\nu} \propto T$. Protoplanetary disks have strong gradients in physical conditions both in radial and vertical direction and thus cannot be taken as at uniform temperature. However, the bulk of the disks mass is close to the midplane, where the physical conditions change less than in the warm molecular layer of the upper disk \citep{vanZadelhoff01,Bruderer12}. Mass tracers such as $^{13}\rm CO$ trace mostly regions close to the midplane. In the 
midplane, the temperature is a weak function of the radius with $T \sim r^{-q}$ with $0 < q < 1$. In the simple power-law case the pivot-point is at radii between 60 and 100 au (see Fig. \ref{pivot_point}) and the temperature at the 
radius of the pivot point varies by less than a factor of two. Thus, the intensity also 
changes by less than a factor of two. This explains the similar intensity at the pivot-point. 
The explanation may also apply to other molecules with emission mostly from regions close to the midplane (e.g. $^{12}\rm CO$ or $\rm C^{18}O$). 

We have compared our results in the simple power-law case and self-similar disk case, to test that one can always retrieve $\gamma$ safely around the same range or line intensity. In Fig. \ref{pivot_point} the $^{13}$CO emission radial profiles are compared with the total line optical depth $\tau$  and the $^{13}$CO column density. The line emission can be used to trace the gas distribution if $\tau<1$ (shaded region in panels a, and d), i.e. for radii larger than those shown by the dotted lines. On the other hand, the $^{13}$CO column density needs to be larger than $10^{15}$ cm$^{-2}$ where the self shielding is effective (shaded region in panels b, and e), i.e. for radii smaller than those shown by the dashed lines. Combining these two requirements one is restricted to a radial range, the "pivot-region", where the $^{13}$CO intensity is always around 6 K km s$^{-1}$, for each value of $\gamma$ and no matter if the underlying $\Sigma_{\rm gas}$ is given by a simple power-law or a self-similar disk model.
For all models analyzed in this work, the minimum and maximum flux values relative to the slope pivot-regions are 3 K km s$^{-1}$ and 8 K km s$^{-1}$. For the Herbig models with $M_{\rm disk}=10^{-3}M_{\odot}$ in particular, if one computes the line optical depth and the column density, as shown in Fig. \ref{pivot_point}, the pivot-point shifts toward larger radii. The slope pivot-region is however still located around the value of 6 K km s$^{-1}$ as for the T Tauri models. This means that, in practice, one would only need to measure the slope of the $^{13}$CO radial intensity over a narrow range of radii close to this contour to retrieve the surface density slope (see shaded region in panels c, and f). The pivot-region will always be very small, as the line intensity that corresponds to N($^{13}$CO)=$10^{15}$ cm$^{-2}$ is very close - within a factor of a few - to that where the line becomes optically thin. The models presented in Fig. \ref{pivot_point} have a total disk mass of $10^{-4} M_{\odot}$ and therefore $^{13}$CO emission lines are still optically thin in the interested region of the disk. If the disk was more massive, for example with $M_{\rm disk}=10^{-2} M_{\odot}$, the same argument described above would be valid at larger radii or for less abundant CO isotopologues, e.g. for C$^{17}$O. Also for C$^{17}$O, one could retrieve $\gamma$ by fitting the radial profile at a given contour at 6 K km s$^{-1}$. In fact, in principle one could put together the surface density profile over the entire radial range by putting together piecemeal the slopes at the pivot points of each of the CO isotopologues, from $^{13}$C$^{18}$O in the innermost part of the disk, to C$^{17}$O, C$^{18}$O, and $^{13}$CO moving outward in the disk, and finally $^{12}$CO itself in the outermost region. If the observed disk is inclined by more than about 30$^{\circ}$, the data should be deprojected in order to apply this method. 

This finding is promising as it relates the observable line intensity radial profile directly with the column density, with no need to know a priori the radial range for the fit of $\gamma$ or disk mass. On the other hand, the observer would need good angular resolution and sensitivity to perform this fit at a given contour. Typically, one would need a resolution of $\sim 0.1"$- 0.2" to determine the slope of the surface density distribution in disks. The peak flux at the slope pivot-point would lie around 10 mJy beam$^{-1}$ km s$^{-1}$ for the $^{13}$CO $J=3-2$ line (and around 3.5 mJy beam$^{-1}$ km s$^{-1}$ for the $^{13}$CO $J=2-1$ transition) at a resolution of 0.1", and 40 mJy beam$^{-1}$ km s$^{-1}$ (and around 14 mJy beam$^{-1}$ km s$^{-1}$ for the $^{13}$CO $J=2-1$ transition) at a resolution of 0.2" for a disk at 150 pc. Requesting a resolution of 0.1" and a S/N $\sim 5$ would make the integration time very long even with ALMA , but increasing the beam size to 0.2" results in sufficient S/N in less than one hour observation, as the data do not need to be spectrally resolved. 

\subsection{A case study: TW Hya}
\label{TWHya}

We test the slope pivot-region fitting on the TW Hya disk observations ($i \sim 6^{\circ}$, $d=60.1$ pc as found by \cite{Bailer-Jones18}) published by \cite{Schwarz16}. More precisely, the C$^{18}$O $J=3-2$ transition has been considered. According to the model presented by \cite{Schwarz16}, the $^{13}$CO $J=3-2$ line is optically thick throughout the entire disk, as expected since TW Hya is a much brighter and more massive disk than those considered in our work. Therefore,  C$^{18}$O is needed in order to constrain $\Sigma_{\rm gas}$, as this is found to be optically thin everywhere. The observation beam size is $\sim 0."5 \times 0."3$ and, at such resolution, the slope pivot-point is expected to be at 0.15 Jy beam$^{-1}$ km s$^{-1}$.

\begin{figure}
   \resizebox{\hsize}{!}
             {\includegraphics[width=1.\textwidth]{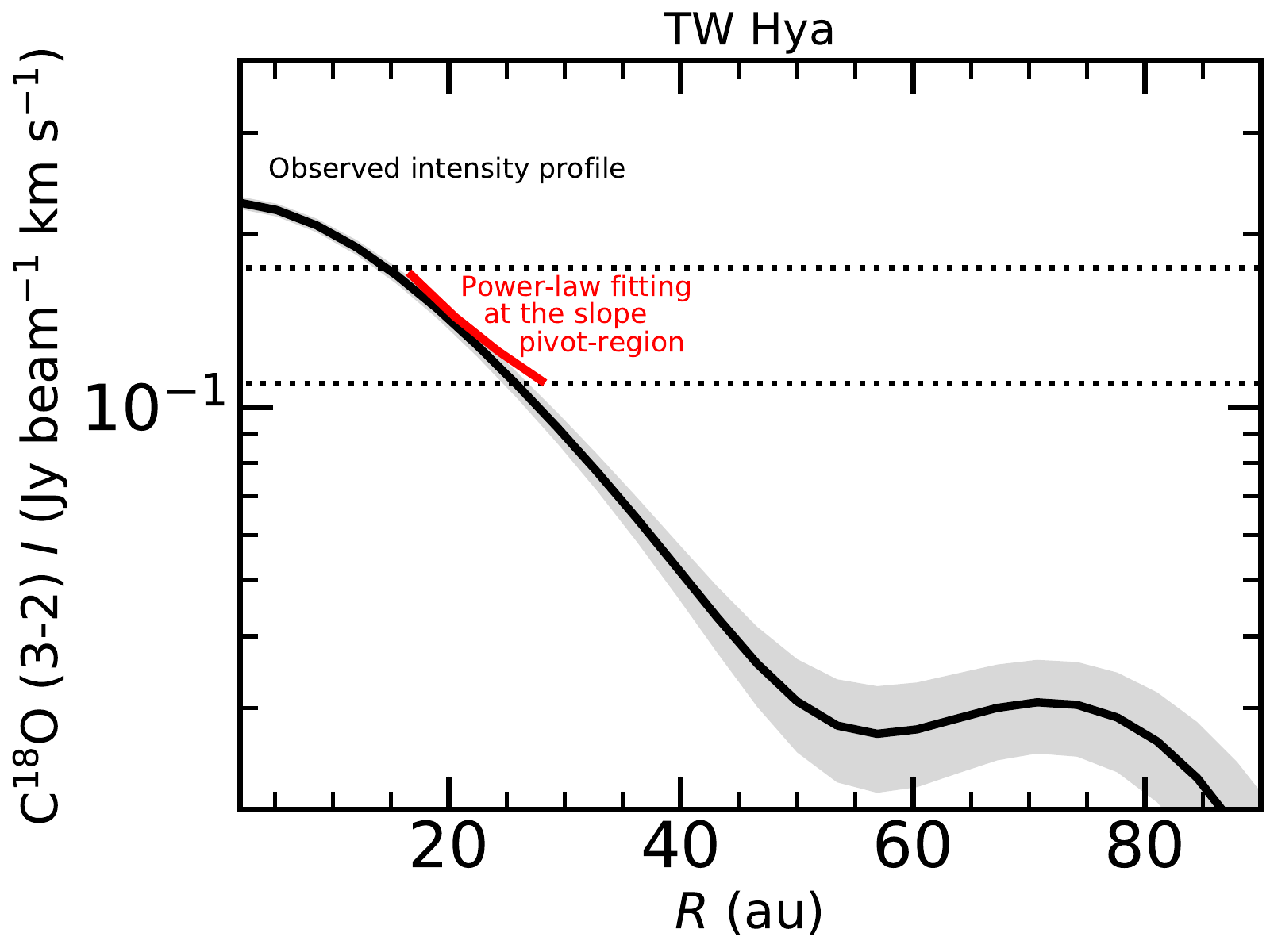}}
      \caption{The radial intensity profile of the C$^{18}$O (3-2) line observed with ALMA \citep{Schwarz16} is shown in black and the 3$\sigma$ level of uncertainty is presented by the shaded region. The red line shows the power-law function used to fit the data at the slope pivot-region.}
       \label{TWHya_fig}
\end{figure}

Fig. \ref{TWHya_fig} shows the the observed C$^{18}$O ($J=3-2$) radial profile of TW Hya in black, with the 3$\sigma$ level of uncertainty shown by the shaded region. If a power-law is fitted to the observed radial profile between $\sim$ 0.1 and 0.2 Jy beam$^{-1}$ km s$^{-1}$, corresponding to radii between $\sim$ 15 and 30 au (see red line in Fig. \ref{TWHya_fig}) the fitted power-law index is $\gamma=0.85$, which is within 15\% of the model value $\gamma=1$ assumed by \cite{Schwarz16} and \cite{Kama16} for the TW Hya disk. With a S/N>20 in the slope pivot region, the error on gamma is negligible, less than $\pm0.05$. The slight mismatch between the data and the power-law profile on the left side of the slope-pivot region is due to the optical thickness of the emission that becomes more important. On the right side of the slope-pivot region, instead, the mismatch is probably due to the fact that C$^{18}$O is being selectively photodissociated. For less massive disks, one should prefer $^{13}$CO over C$^{18}$O to trace $\Sigma_{\rm gas}$, as the first is less affected by isotope-selective photodissociation compared with the latter \citep[as shown by][]{Miotello14}.

High angular resolution and sensitivity are however needed to apply this method, but ALMA can achieve them. In the particular case of TW Hya, where our method could be applied, the rms was 12 mJy beam$^{-1}$ in a 0.1 km s$^{-1}$ channel. 

\section{Discussion}
\label{discussion}

The results presented in Sec. \ref{results} indicate that constraining $\Sigma_{\rm gas}$ from spatially resolved observations of $^{13}$CO is a challenging task. However, if the underlying surface density distribution is a power-law, it is possible to constrain its steepness by fitting the emission coming from the correct portion of the disk, in particular at large enough distances from the star that the observed line is optically thin, but at small enough radii that photo-dissociation and freeze-out are not too significant. Moreover, in these intermediate regions, the radial gradient of gas temperature is small, and thus does not affect the gradient of the radial intensity profile of the $^{13}$CO line. From the models shown in this paper, it is possible to retrieve the dependence of the surface density with radius with good accuracy ($20\%$ on the power-law exponent) especially if one has an estimate of the $^{13}$CO total mass or if one has high enough resolution data to fit around the pivot-point-region at 6 K km s$^{-1}$.

However, in the approach presented in Sec. \ref{fit} and in Sec. \ref{sec:exp}, there is the very strong assumption that the surface density is either a power law function of the radius or it is given by the self-similar model. However, the actual underlying surface density in real protoplanetary disks is of course not known. Moreover, the main difficulty in the fitting is due to the relatively narrow radial range where $^{13}$CO traces well the underlying surface density, with this radial range limited by optical depth effects in the inner disk and inefficient self-shielding in the outer regions. Determining additional free parameters, such as the characteristic radius of the self-similar profile, leads to a large degeneracy between the power-law index and the new parameters. To overcome the issue of $^{13}$CO being optically thick at small radii, one possibility is to probe $\Sigma_{\rm gas}$ in the inner disk by observing very optically thin isotopologues, such as C$^{17}$O and $^{13}$C$^{18}$O. The latter was successfully detected with ALMA in the closest and best studied protoplanetary disk, TW Hya, by \cite{Zhang17}. The inconveniences of this emission line are its faintness combined with the possibility that the continuum becomes optically thick shielding $^{13}$C$^{18}$O emission at very small radii.
Moreover the gas temperature structure has a strong impact on the line emission in the inner disk, therefore possibly independent constraints of the thermal structure would be needed.

An much simpler approach is to fit the power-law index $\gamma$ at the slope-pivot-region as explained in Sec. \ref{slope-pivot-point}. This allows a reliable estimate of $\gamma$ both in the case of a simple power-law and self similar disk, by only fitting the intensity profile at the right intensity contour. The case study presented in Sec. \ref{TWHya} shows the applicability of this method to resolved observations of disks with ALMA.

An additional caveat is volatile carbon depletion, a process that may be happening in a large fraction of protoplanetary disks \citep{Favre13,Kama16,Schwarz16,Miotello17}. In our models, the volatile carbon abundance is assumed to be high and constant throughout the disk. If carbon depletion takes place, but it is constant throughout the disk especially around the pivot-region, this should not add major complications to retrieving the surface-density profile. However, the nature of this depletion process is not yet known, but it is possible that carbon is sequestered from CO into CO$_2$ and more complex ice species in the outer disk, and then drifted inward following the large dust grains. If this is true, one would expect a radially dependent decrease of CO abundances in the outer disk, on top of the reduction due to inefficient self-shielding and freeze-out. As the ices reach the inner disk then carbon should be liberated into gas phase and quickly return into CO, which would then present an increased emission \citep{Du15}. This would introduce a new degree of degeneration in the employment of CO isotopologues as tracers of the disk surface density distribution. Interestingly, the expected enhancement of $^{13}$C$^{18}$O in the inner disk of TW Hya has not been found by \cite{Zhang17}, showing that much is still to be understood about volatile depletion in protoplanetary disks.

\section{Summary and conclusion}
In this work we have addressed the issue of determining the gas surface density distribution in protoplanetary disks by using resolved high signal-to-noise observations of $^{13}$CO. Simulated $^{13}$CO intensity radial profiles have been produced using the physical-chemical code DALI \citep{Bruderer12,Bruderer13}, with two different input surface density profiles: a simple power-law with radius, and the self-similar solution given by viscously evolving disk theory. By comparing the input $\Sigma_{\rm gas}$ profiles with the output intensity profiles one always finds the following:
\begin{itemize} 
\item $^{13}$CO emission follows the slope of $\Sigma_{\rm gas}(R)$, but only in a specific disk region. For very small radii the low-$J$ $^{13}$CO lines become optically thick and underestimate the surface density, while in the outer disk this happens because the $^{13}$CO column density drops below $10^{15} \, \rm cm^{-2}$ and self-shielding becomes inefficient. 
\item When fitting a power-law to the simulated intensity profiles, it is possible to recover the input power-law index $\gamma_{\rm input}$ only by performing the fit over the appropriate radial range. This holds in particular for simple power-law disk models, where $\gamma_{\rm input}$ can be retrieved within 20$\%$ uncertainty between 30--100 au, even when the profiles are convolved with a 0.2" beam.
\item In the self-similar case it is not always possible to reliably retrieve $\gamma_{\rm input}$ by fitting a self-similar model to the intensity profile. $R_{\rm c}$ can instead be always retrieved within 30 \%.
\item Fitting the power-law index $\gamma$ in a narrow range around the slope-pivot-region of the intensity profile allows a reliable estimate of $\gamma$ both in the case of a simple power-law and self-similar disk. The slope-pivot-region is always located around 6 K km s$^{-1}$. Application of such a method is shown in the case study of the TW Hya disk.
\item If carbon depletion were constant throughout the disk, this would not introduce an additional uncertainty in the employment of CO isotopologues as tracers of the disk surface density distribution. 
\end{itemize}
$^{13}$C$^{18}$O may be a better tracer of $\Sigma_{\rm gas}(R)$ in the inner regions for massive disks, circumventing the $^{13}$CO optical depth issue, as suggested by \cite{Zhang17}. For lower mass disks, C$^{17}$O and C$^{18}$O may be more appropriate. However, both dust optical depth and gas temperature may limit the analysis. Thus, combining observations of optically thin tracers with $^{13}$CO may be the best option after-all.

\section*{Acknowledgements}

The authors thank the referee J. P. Williams, S. Andrews, L. Testi, I. Pascucci, and I. Kamp for the comments which helped to improve the paper, and E. Bergin, and K. Schwarz for sharing their data. Astrochemistry in Leiden is supported by the Netherlands Research
School for Astronomy (NOVA), by a Royal Netherlands Academy of Arts
and Sciences (KNAW) professor prize, and by the European Union A-ERC
grant 291141 CHEMPLAN. AM acknowledges an ESO Fellowship.


\clearpage
\begin{appendix}
\section{Additional figures}
\label{lower_mass}

Here some ancillary figures are reported. 

\begin{figure*}
   \resizebox{\hsize}{!}
             {\includegraphics[width=1.\textwidth]{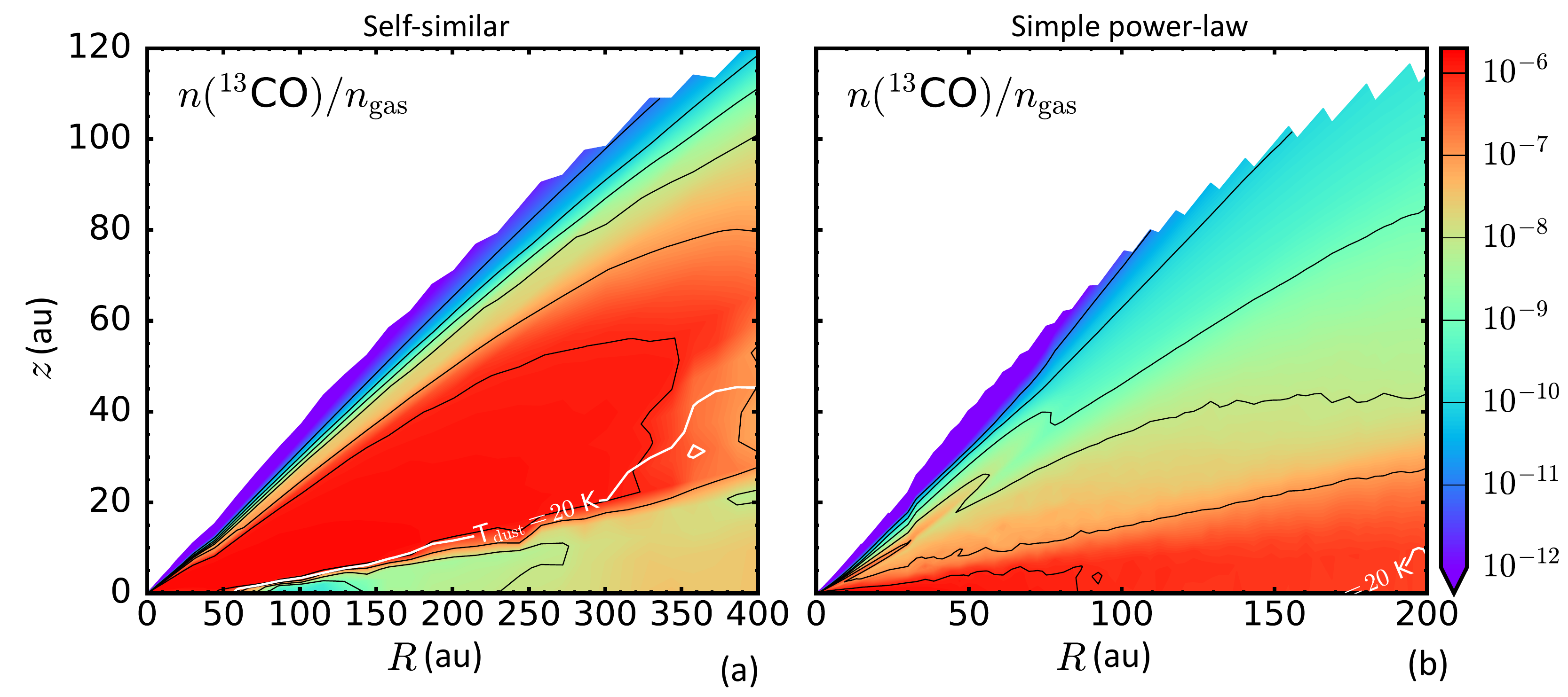}}
      \caption{Abundance of $^{13}$CO in one quadrant the disk for two representative models. Panel (a) shows the self-similar disk model ($R_{\rm c}=200$ au, $M_{\rm disk}=10^{-3}M_{\odot}$ and $\gamma=1$); panel (b) shows the simple power-law disk model ($R_{\rm out}=200$ au, $M_{\rm disk}=10^{-4}M_{\odot}$ and $\gamma=1$). The white contours indicates the $T_{\rm dust}$=20 K surface below which CO freeze-out becomes important.}
       \label{n13CO}
\end{figure*}

\begin{figure*}[b]
   \resizebox{0.6\hsize}{!}
             {\includegraphics[width=\textwidth]{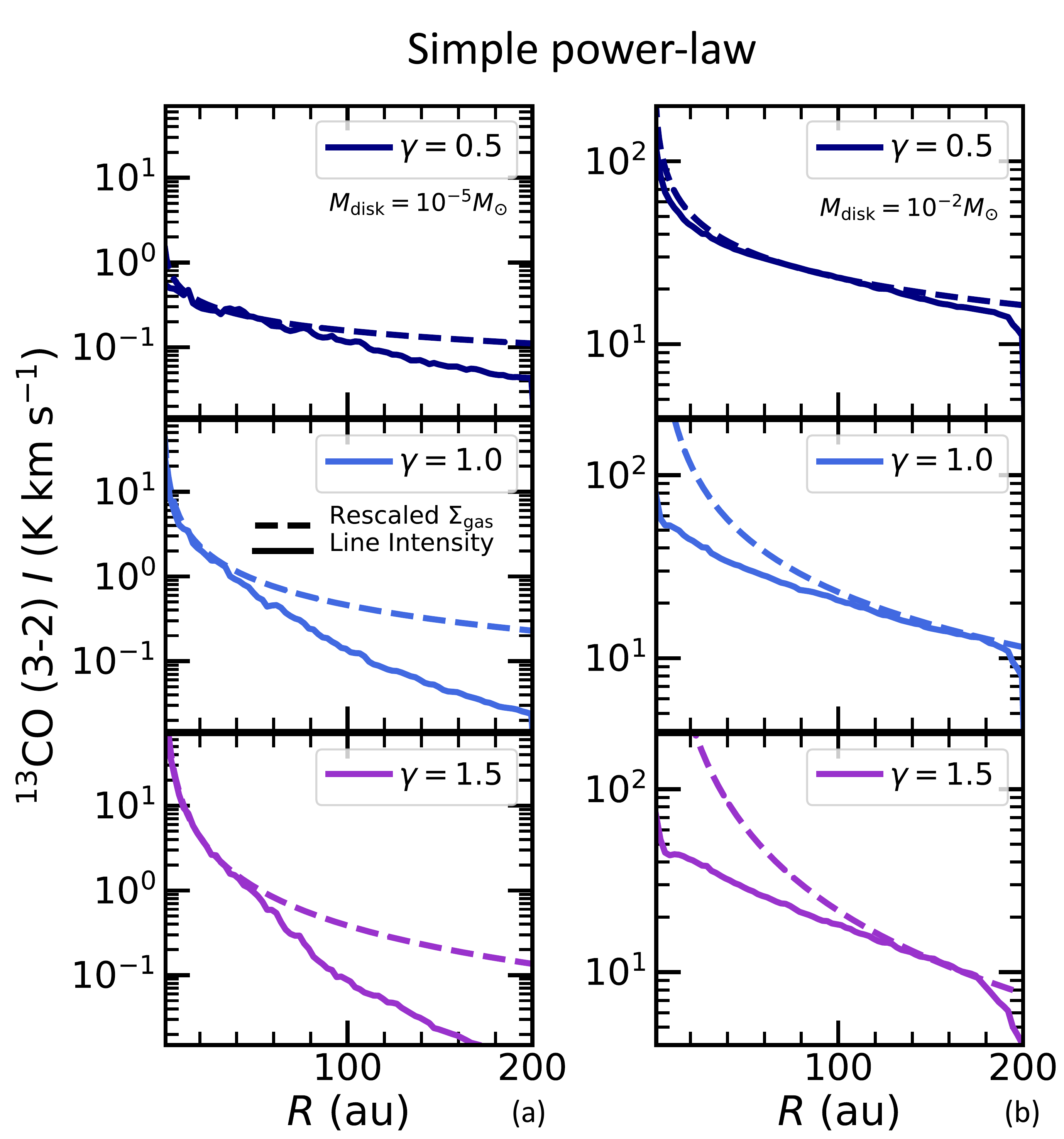}}
\centering
      \caption{$^{13}$CO line intensity radial profiles (solid lines) obtained with input surface density distribution $\Sigma_{\rm gas}$ (dashed lines) given by the simple power-law disk models (see Eq. \ref{sigma_eq}). The model parameters are $M_{\rm disk}=10^{-5}M_{\odot}$ (panel a) and $M_{\rm disk}=10^{-2}M_{\odot}$ (panel b), $\gamma=0.8 ,1,1.5$ shown in dark blue, light blue, and purple respectively (top, middle, and bottom panels).}
       \label{prof_masses}
\end{figure*}

\begin{figure*}[h]
   \resizebox{\hsize}{!}
             {\includegraphics[width=1.\textwidth]{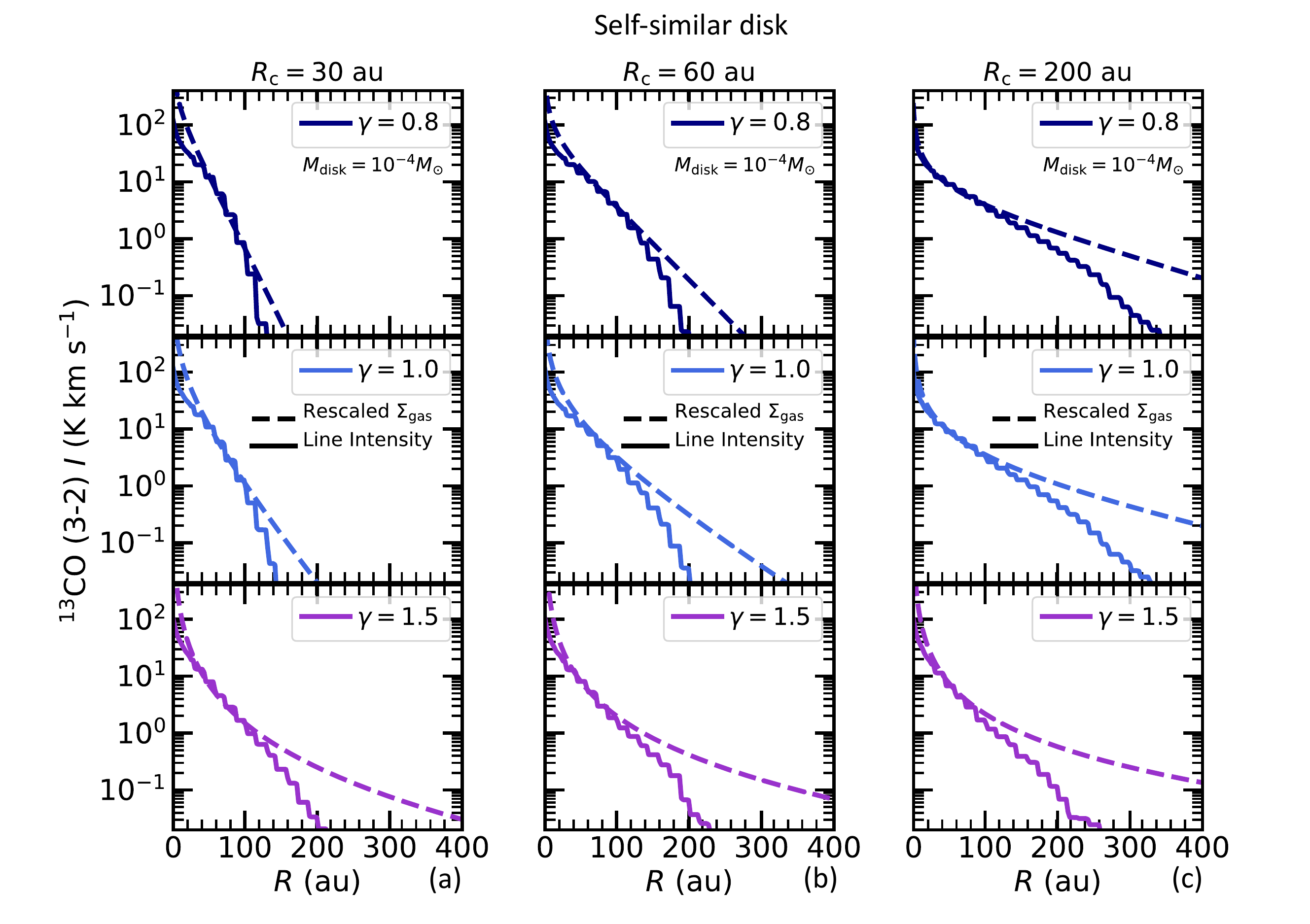}}
      \caption{$^{13}$CO line intensity radial profiles (solid lines) obtained with input surface density distribution $\Sigma_{\rm gas}$ (dashed lines) given by the self-similar disk model (see Eq. \ref{sigma_eq}). The model parameters are $M_{\rm disk}=10^{-4}M_{\odot}$ and $\gamma=0.8 ,1,1.5$ shown in dark blue, light blue, and purple respectively (top, middle, and bottom panels). Model with $R_{\rm c}=30, 60, 200$ au are presented in panel a, b, and c respectively.}
       \label{exp_low_mass}
\end{figure*}

\end{appendix}
\end{document}